\documentclass[aps, pre, secnum, twocolumn]{revtex4-2} 

\usepackage{graphicx}
\usepackage{amsmath}
\usepackage{algorithm}
\usepackage{algorithmicx}
\usepackage{algpseudocode} 
\usepackage{setspace} 
\usepackage{caption}      
\usepackage{float}         
\usepackage{xcolor}  

\newcommand{\ud}{\mathop{}\!\mathrm{d}}

\begin{document}


\title{From informal markets to Limit Order Book dynamics: a mean field connection}


\author{Navarro-Rubio, A.}
\affiliation{Group of Complex Systems and Statistical Mechanics, Physics Faculty, University of Havana}

\author{Lage-Castellanos, A.}
\email[]{ale.lage@gmail.com}
\affiliation{Group of Complex Systems and Statistical Mechanics, Physics Faculty, University of Havana}


\date{\today}

\begin{abstract}
    We propose a unified mean-field framework that bridges the dynamics of informal financial markets and formal markets governed by Limit Order Books (LOBs). Both settings are modeled as interacting particle systems on a 1D price lattice, with temporal evolution described by master equations that account for new entries, cancellations, and executions. The key insight is the introduction of a preferential interaction parameter $\Psi$, which modulates the likelihood of transactions based on price compatibility: when $\Psi=0$, interactions are random and uncoordinated, reproducing the structure of informal markets; as $\Psi\to \infty$, only optimal (most mutually attractive) trades occur, recovering LOB-like dynamics. A grand-canonical interpretation is used to identify effective thermodynamic quantities—such as interaction energy and price-dependent chemical potentials—that underlie both systems.  Most results are validated through numerical integration and simulations, although an analytical solution is shown to exist at least for the symmetric stationary case of the  informal market.  
\end{abstract}


\maketitle

\section{\label{intro}Introduction}
	
	    Financial markets vary widely in their degree of institutionalization, ranging from highly regulated exchanges governed by transparent rules to informal, unregulated settings that emerge in response to policy distortions or institutional voids. While the dynamics of formal markets—particularly those operating via Limit Order Books (LOBs)—have been extensively studied in econophysics \cite{bouchaudTradesQuotesPrices2018,slaninaEssentialsEconophysicsModelling2013}, informal markets remain comparatively underexplored, largely due to data scarcity and their context-specific nature. Yet, understanding both is essential: informal markets often dominate in developing economies and can offer insights into market behavior under minimal institutional scaffolding.

    From a modeling perspective, LOBs are typically represented as discrete price lattices where buy and sell orders arrive, cancel, and execute according to stochastic rules. Zero-intelligence agent-based models—such as the Santa Fe model—have successfully reproduced key empirical features (e.g., order book shape, spread dynamics) using minimal behavioral assumptions \cite{farmerPredictivePowerZero2005,StatisticalTheoryContinuous2003}. In contrast, recent work on the Cuban informal foreign exchange market has modeled transactions as uncoordinated interactions between agents posting intentions on social media, with prices anchored to a publicly announced reference value \cite{delaolivacuelloModeloGasPara2024,figalLookingInformalCurrency2025}. Although both approaches employ particle-gas analogies and mean-field master equations, their formulations differ in execution logic, limiting direct comparison.

    In this paper, we reconcile these two strands by developing a generalized mean-field framework that encompasses both market types as limiting cases of a single dynamical system. The core innovation is a preferential interaction mechanism, controlled by a parameter $\Psi$, which weights the probability of a transaction by the mutual attractiveness of the participating prices (i.e., how close they are to optimal execution). When $\Psi=0$, all compatible trades are equally likely—a regime that reproduces the random matching characteristic of informal markets. In the limit $\Psi\to \infty$, only the most favorable matches occur, mimicking the price-time priority of LOBs. Thus, $\Psi$ acts as a tunable selectivity parameter, not a true thermodynamic temperature, but one that governs the degree of coordination in the system.

    We derive master equations for both limits and show that they emerge from a common structure. These are validated against stochastic simulations (agent-based and Gillespie-based) and numerical integration, with excellent agreement in both transient dynamics and stationary profiles. For the informal limit, we provide an exact analytical solution of the stationary state under symmetric Gaussian assumptions. Furthermore, we interpret the stationary distributions through a grand-canonical lens, defining effective interaction energy and price-dependent chemical potentials that encode entry, cancellation, and execution balances.

    This unified perspective demonstrates that the distinction between informal and formal markets is not categorical but continuous and parameter-driven. It enables techniques developed for LOB modeling to be adapted to informal contexts—and vice versa—while offering a minimal statistical-mechanical foundation for market diversity.

	

	\subsection{\label{lobs intro}Limit Order Books}
	
	\begin{figure}
			\centering
			\includegraphics[width=\columnwidth, keepaspectratio]{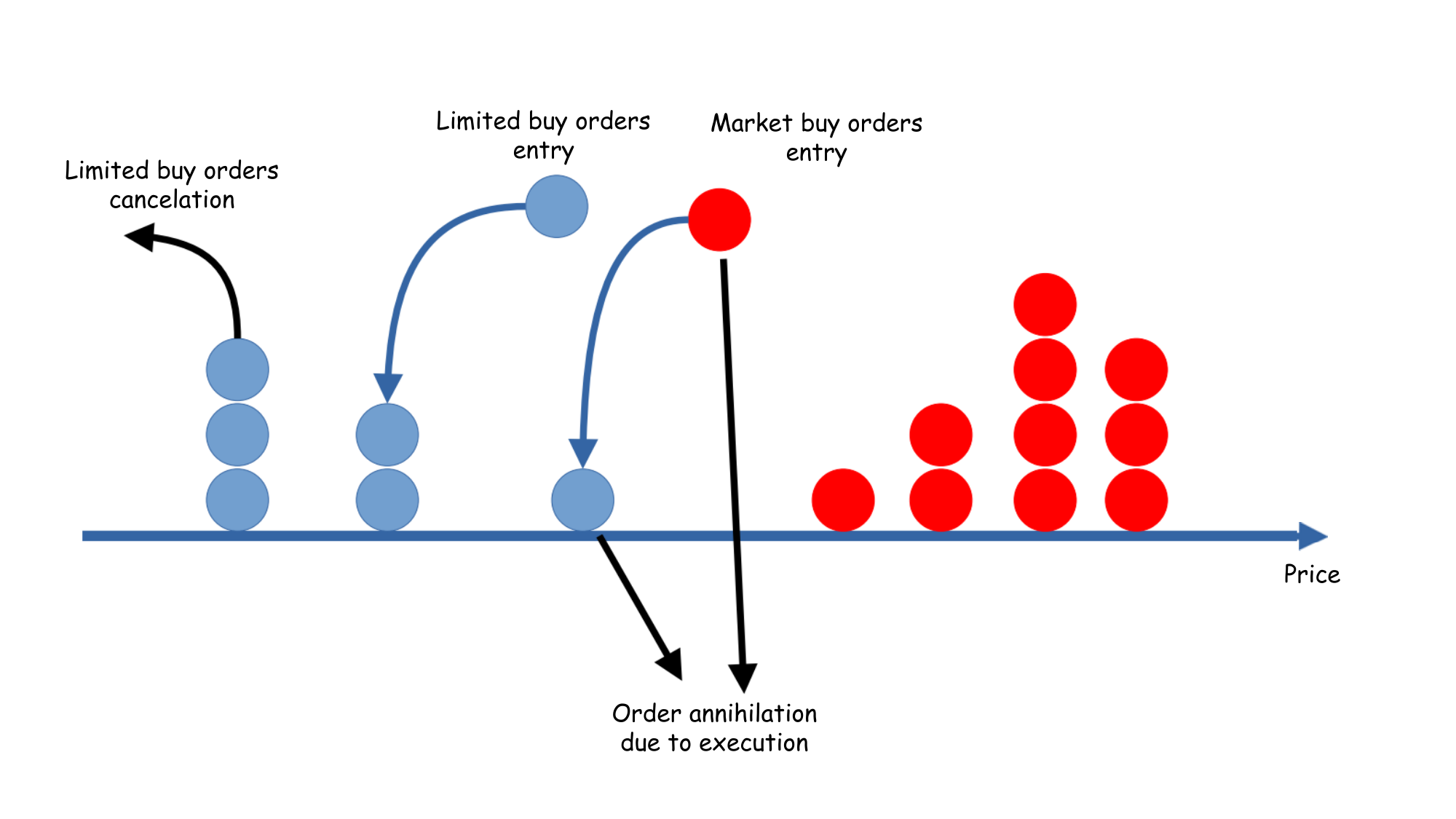}
			\caption{Illustration of the processes that define order flow by price range. Entries are defined as the arrival of limit or market orders, while exits occur due to spontaneous voluntary cancellations and order executions.}
			\label{fig:procesos LOB}
		\end{figure}

		A Limit Order Book constitutes today’s most common infrastructure of \textit{lit} financial markets, where buy and sell orders are visible to all participants \cite{bouchaudTradesQuotesPrices2018}. The LOB electronically organizes such orders for a specific asset according to price and volume, structured into two sides: bid and ask. The best bid price corresponds to the highest value offered by a buyer, while the best ask price is the lowest value accepted by a seller; the difference between the two is called the \textit{spread}, and its average defines the mid-price. Limit orders, those in which a desired execution price is specified, provide liquidity and are executed ensuring time priority; while market orders consume liquidity by executing immediately against the best available price (see Fig. \ref{fig:procesos LOB}).
		
		The LOB operates on a discrete grid of prices and volumes, determined by the price \textit{tick} and the minimum lot size. Its state is defined by the set of active limit orders at a given moment. A market order is matched with limit orders on the opposite side, starting from the best price and moving toward less attractive price levels until its total volume is satisfied.
		
		Modeling the LOB faces challenges associated with the high dimensionality of the state space and the complexity of individual decisions \cite{gouldLimitOrderBooks2013}. From the Econophysics perspective, agent-based models (ABM) have been proposed, mainly of two types: zero-intelligence (ZI) models and rational-agent models. ZI models assume simple stochastic rules without complex strategies, aiming to explore to what extent the observed complexity can emerge from random mechanisms under minimal constraints \cite{slaninaEssentialsEconophysicsModelling2013}.
		
		In a typical ZI formulation, the LOB is represented as a one-dimensional price lattice where orders are introduced, canceled, or executed according to probabilistic dynamics \cite{abergelEconophysicsOrderdrivenMarkets2011}. A representative example is the Santa Fe model \cite{farmerPredictivePowerZero2005,StatisticalTheoryContinuous2003,danielsQuantitativeModelPrice2003}, in which unit-volume orders arrive and are canceled following Poisson processes, and are executed when matched with a counterpart at the most attractive price level.
		
	\subsection{\label{informal intro} Informal markets}
		
		Informal financial markets, also referred to as fragmented, parallel, or black markets, emerge in developing economies as a response to government policies that create mismatches between supply and demand, especially under exchange controls and fixed exchange rate regimes \cite{agenorParallelCurrencyMarkets1990}. Unlike formal markets, they lack regulated legal frameworks and exhibit distinctive features such as low transparency, unregistered transactions, absence of legal protection, predominance of cash, and low entry barriers \cite{delaolivacuelloModeloGasPara2024}.
		
\begin{figure}
			\includegraphics[width=\columnwidth]{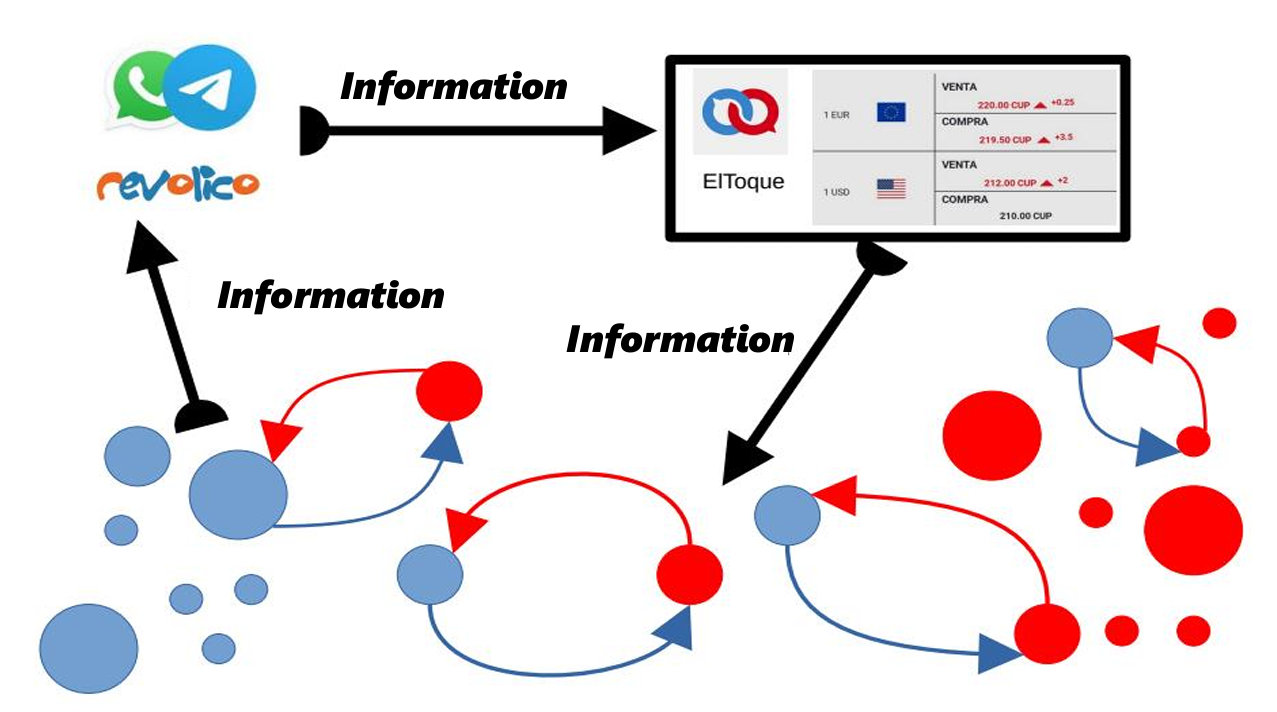}
			\caption{Information flow in the Cuban informal foreign exchange market. Reference price is established daily by elTOQUE platform using messages containing buy/order intentions gathered from social media and digital trade platforms.}
			\label{fig:diagrama del toque}
		\end{figure}

		Information about these markets largely comes from indirect observations, such as prices published on digital platforms. In the case of the Cuban foreign exchange market, Figure \ref{fig:diagrama del toque} shows the mechanism through which the platform elTOQUE calculates a daily reference price based on advertisements posted on social networks, influencing future buying and selling intentions \cite{delaolivacuelloModeloGasPara2024}. However, such a reference price does not always reflect the actual transaction prices, which are often executed in private spaces without institutional oversight \cite{vidalUsingAIInformal2024}.
		
		From the perspective of Econophysics, the study of these markets remains scarce due to their limited presence in developed economies. A relevant contribution is the research carried out by the Center for Complex Systems at the Faculty of Physics, University of Havana, focused on the Cuban informal foreign exchange market \cite{delaolivacuelloModeloGasPara2024,figalLookingInformalCurrency2025,garcia-figalEfficiencyInformalCurrency2025}. In this model, agents represent individuals with foreign currency exchange needs, with prices determined around a reference value based on public advertisements. The model includes processes such as the random arrival of exchange needs, price drift due to the accumulation of orders not yet executed, and agent matching to complete transactions.
		
		Although this approach allows for describing the phenomenology of the Cuban informal market, the implemented rules for order flows and executions depend on hypotheses specific to the case study, which makes it difficult to generalize to other contexts.
	
\section{Probabilistic Model of the LOB}

	LOBs currently constitute the main operating mechanism of global stock markets. Consequently, they have attracted significant interest for decades within the scientific community, particularly among researchers in the field of Econophysics, who have devoted considerable effort to modeling and analyzing the dynamics that characterize them. In this section, a probabilistic analysis of LOB dynamics is proposed, aiming to derive a master equation that describes the profiles of buy and sell orders per price level.
	
	\subsection{Master Equation}
	\label{sub: master eq LOB Lage}

		To analyze the dynamics of the LOB and facilitate comparison with the behavior of informal markets, a model inspired by the previously mentioned Santa Fe framework is proposed, in which buy and sell orders participate in three processes: entry, cancellation, and execution. Figure \ref{fig:procesos LOB} illustrates the inflow and outflow of orders that define this dynamics.

		Order entry is modeled as a Poisson process with rate $\phi$, with prices $p$ distributed according to a probability density dependent on the reference price $s(t)$. Cancellation occurs at a fixed rate $\Delta$, proportionally to the number of orders at each level. Executions happen when market orders find limit counterparts with favorable prices, assuming all orders are of unit size.
		
		A key difference from the discrete treatment offered by the Santa Fe model lies in the adoption of a single Poisson process that unifies the arrival of limit and market orders, which simplifies the dynamics and allows for a more tractable continuous description.
	
		The temporal evolution of the number of orders per price level is formalized through a master equation, composed of entry, cancellation, and execution terms:
		\begin{equation}\label{ec:generical LOB master eq}
			\frac{\ud n(p,t)}{\ud t} = T_{\text{entry}} - T_{\text{cancellation}} - T_{\text{execution}}
		\end{equation}
		
		A probabilistic analysis leads to the coupled system of equations in \eqref{ec: ecuacion maestra dinamica LOB Lage} for sell orders $n_a(p,t)$ and buy orders $n_b(p,t)$ (for brevity $n_a(p), n_b(p)$):
\begin{widetext}
		\begin{equation}
			\small
			\begin{aligned}
				\frac{\ud n_a(p)}{\ud t}
				&= \phi_a P_a(s(t),p) \left[ \frac{1}{N_b(t)} \int_0^p n_b(p',t) \ud p' \right]^{N_b(t)}
				- \Delta_a n_a(p)
				 - n_a(p) \left[ \frac{1}{N_a(t)} \int_p^\infty n_a(p',t) \ud p' \right]^{N_a(t) - 1} \int_p^\infty \phi_b P_b(s(t),p') \ud p' \\
				\frac{\ud n_b(p)}{\ud t}
				&= \phi_b P_b(s(t),p) \left[ \frac{1}{N_a(t)} \int_p^\infty n_a(p',t) \ud p' \right]^{N_a(t)}
				- \Delta_b n_b(p)
				- n_b(p) \left[ \frac{1}{N_b(t)} \int_0^p n_b(p',t) \ud p' \right]^{N_b(t) - 1}
				\int_0^p \phi_a P_a(s(t),p') \ud p'
				\label{ec: ecuacion maestra dinamica LOB Lage}
			\end{aligned}
		\end{equation}
\end{widetext}	where each term reflects a specific process:
		\begin{itemize}
		\item The first term represents the arrival of new orders, weighted by the probability that they are not immediately executed.
		\item The second describes the spontaneous cancellation of active orders with probability $\Delta$,
		\item The third quantifies executions due to incoming market orders interacting with competitive limit counterparts.
		\end{itemize}
		The incoming distribution of orders $P_a(s,p)$ and $P_b(s,p)$ are left in full generality.
		
		\subsection{Simulation and Comparison}\label{sub:Simulacion LOB}
		
			The dynamics of the LOB were simulated using an ABM implemented in \textit{Python}, considering symmetric parameters for $n_a(p,t)$ and $n_b(p,t)$, as shown in Algorithm \ref{alg:LOB}. The price arrival distribution was defined as $P(s(t),\sigma^2,p)$, a Gaussian centered at the reference price fixed in time $s=100$ with unit variance. The price space was discretized into 500 \textit{bins} within the interval $[100 - 5\sigma,\ 100 + 5\sigma]$, with finite step $dp$.
			
			Although empirically is has been shown that the distribution of orders as a function of distance $\Delta_p$ from the reference price follows a power law \cite{bouchaudTradesQuotesPrices2018}, the proposed model allows for the incorporation of such distributions. Similarly, an extension allowing the update of $s(t)$ could enable the exploration of recovering empirical phenomena such as heavy tails in returns \cite{EmpiricalPropertiesAsset2001} or volatility clustering \cite{andersenDistributionRealizedStock2001}, although such studies are beyond the scope of this work.
			
			Orders enter as a Poisson process with rate $\phi = 20$ and are canceled at a constant rate $\Delta = 0.01$. Classification between limit and market orders is determined according to the incoming price relative to the best available on the opposite side. The values of $\phi$ and $\Delta$ were chosen for convenience, as their ratio controls the total number of orders in the steady state and the system’s stabilization time, balancing size and computational efficiency.
			
			Ten independent simulations of duration $T = 1500$ were performed, averaging results and standard deviations. To obtain the stationary profiles, in each simulation the last 750 histograms after reaching equilibrium were averaged and then averaged across simulations. Simultaneously, the master equation \eqref{ec: ecuacion maestra dinamica LOB Lage} was numerically integrated with step $dt=0.1$ using the Runge-Kutta 5(4) method from the \textit{SciPy} package \cite{virtanenSciPy10Fundamental2020}, using 1000 \textit{bins} and the same parameters as in the simulation.
			
\begin{figure}
				\centering
				\includegraphics[width=\columnwidth, keepaspectratio]{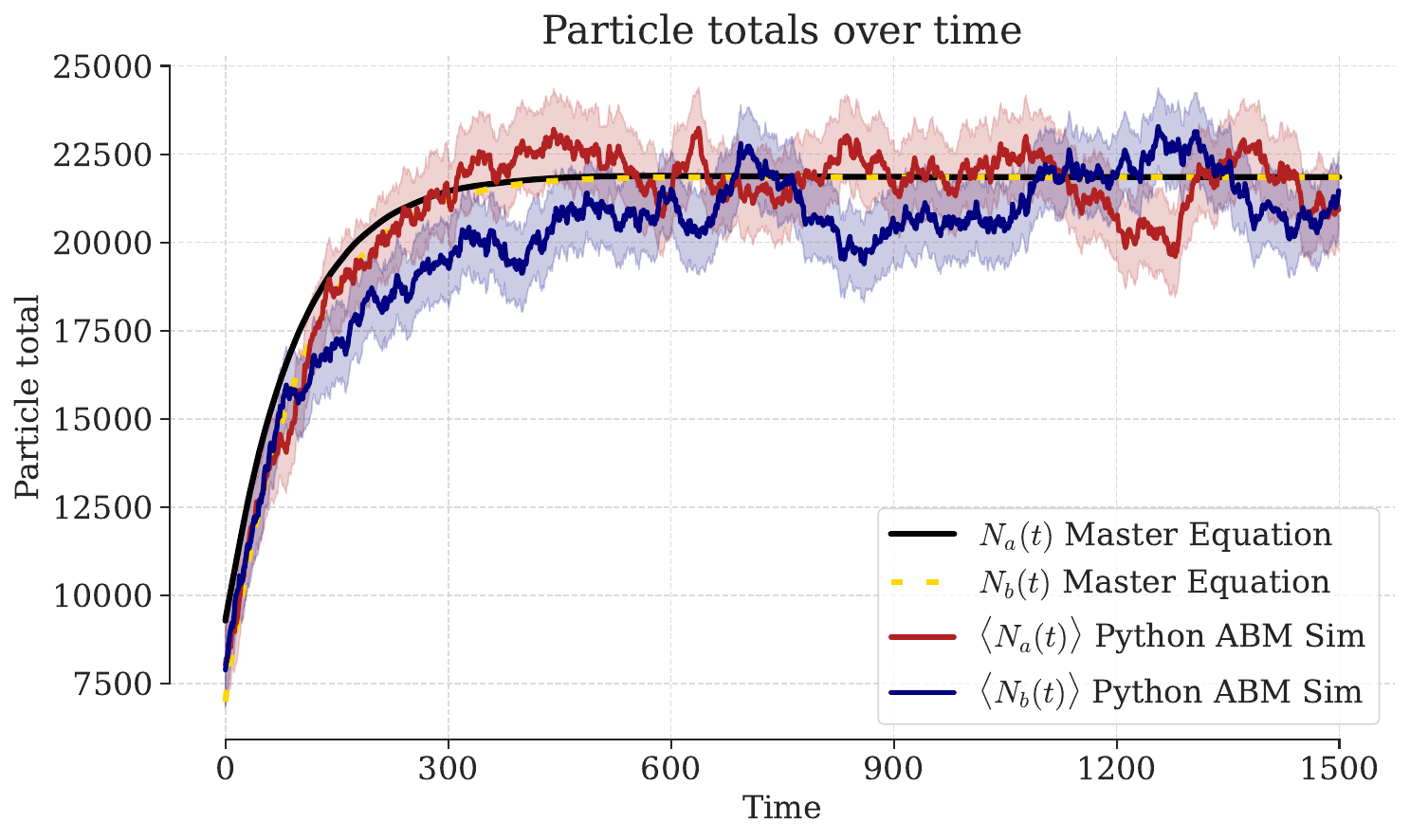}
				\caption{Average total buy and sell orders as a function of time, obtained through stochastic simulation and numerical integration of the master equation \ref{ec: ecuacion maestra dinamica LOB Lage}. The error bands corresponding to one standard deviation are also shown.}
				\label{fig:totales de particulas LOB}
			\end{figure}

\begin{figure*}
				\centering
				\includegraphics[width=0.8\textwidth, keepaspectratio]{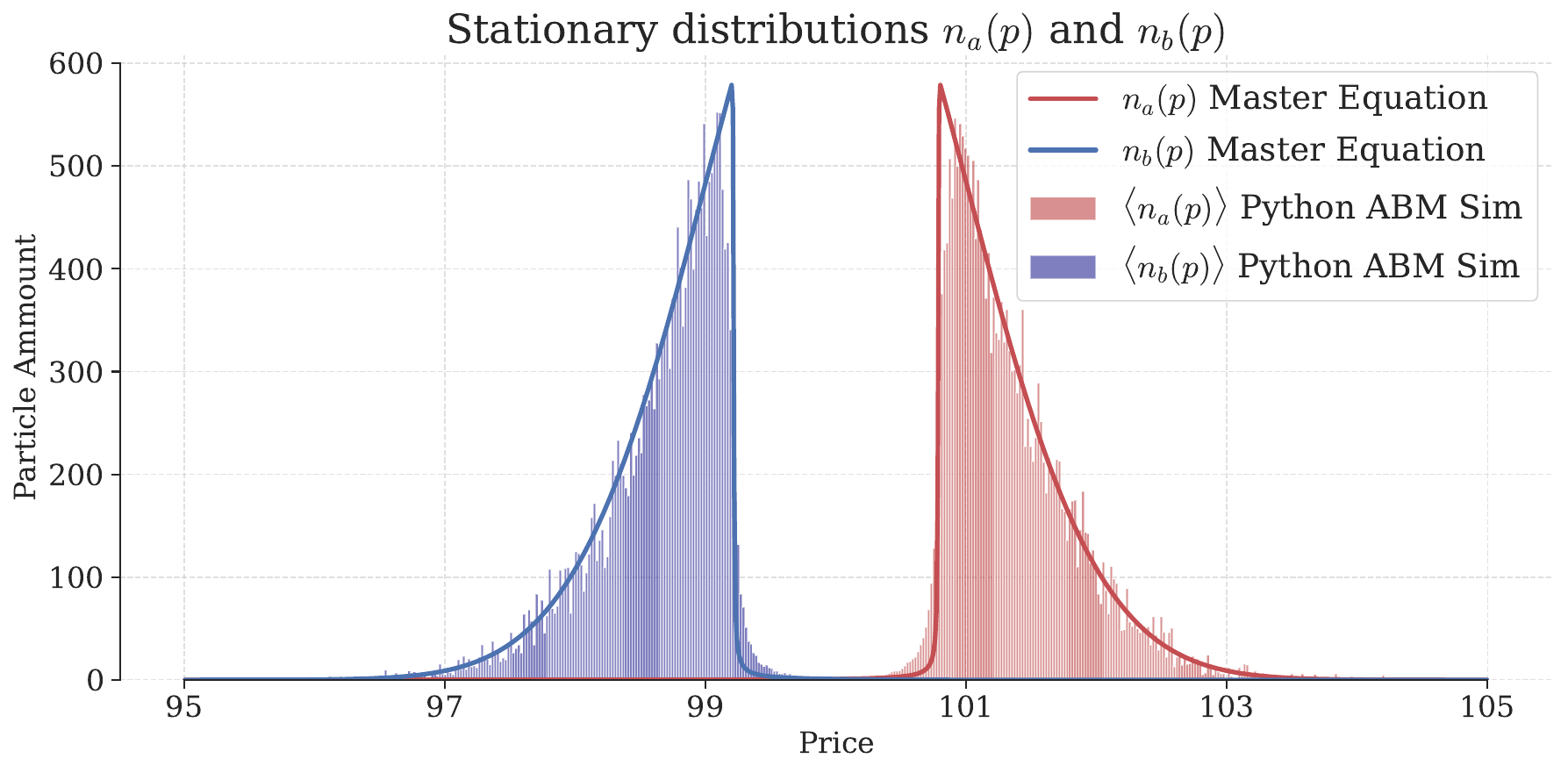}
				\caption{Averaged stationary profiles of $n_a(p)$ and $n_b(p)$ obtained through stochastic simulation and integration of the master equation \ref{ec: ecuacion maestra dinamica LOB Lage}.}
				\label{fig:perfiles na y nb LOB}
			\end{figure*}

			The results shown in Figures \ref{fig:totales de particulas LOB} and \ref{fig:perfiles na y nb LOB} display excellent agreement between the stochastic simulations and numerical integration, both in the temporal evolution of total orders and in the stationary profiles of $n_a(p)$ and $n_b(p)$. Although the master equation does not reproduce stochastic fluctuations, it accurately captures the average dynamics of the system.
			
\section{Informal Gas Model}

	Following the approach proposed by \cite{delaolivacuelloModeloGasPara2024}, in which the dynamics of an informal foreign exchange market are modeled using mean-field master equations describing the profiles $N_a(p,q,t)$ and $N_b(p,q,t)$ which represent the distributions of sell and buy orders, respectively, as functions of price, volume, and time; we aim to re-derive these equations. The objective is to reformulate the underlying processes so that their description more closely resembles that used for the Limit Order Book, which will facilitate, in later stages, the identification of possible theoretical connections between both dynamics.

	\subsection{Description}

		We propose a model in which agents submit, at each time $t$, buy or sell orders of a certain quantity $q$ at a price $p$. The parameters $(p,q)$ are distributed according to a joint probability density function $P(s(t),p,q)$, conditioned on a reference price $s(t)$. The state space $(p,q)$ is discretized into two grids with finite increments $dp$ and $dq$, which define the system’s resolution.
		
		Order entry is modeled as Poisson processes with rates $\phi_a$ and $\phi_b$ for sell and buy orders, respectively, while voluntary cancellation is introduced with a constant probability for each order, governed by parameters $\Delta_a$ and $\Delta_b$.
		
		Executions are governed by interaction rates $\kappa_a$ and $\kappa_b$, proportional to the quantity of opposite-type orders in the system at time $t$. A transaction occurs only if a buy order for $q_b$ units at price $p_b$ interacts with a sell order for $q_a$ units at price $p_a$, under the condition $p_b \geq p_a$. If $q_a = q_b$, both orders are fulfilled and removed from the book; in cases where $q_a > q_b$ or $q_a < q_b$, the smaller-volume order is exhausted and the residual order remains in the market until fully executed.
		
\begin{figure}
			\centering
			\includegraphics[width=\columnwidth, keepaspectratio]{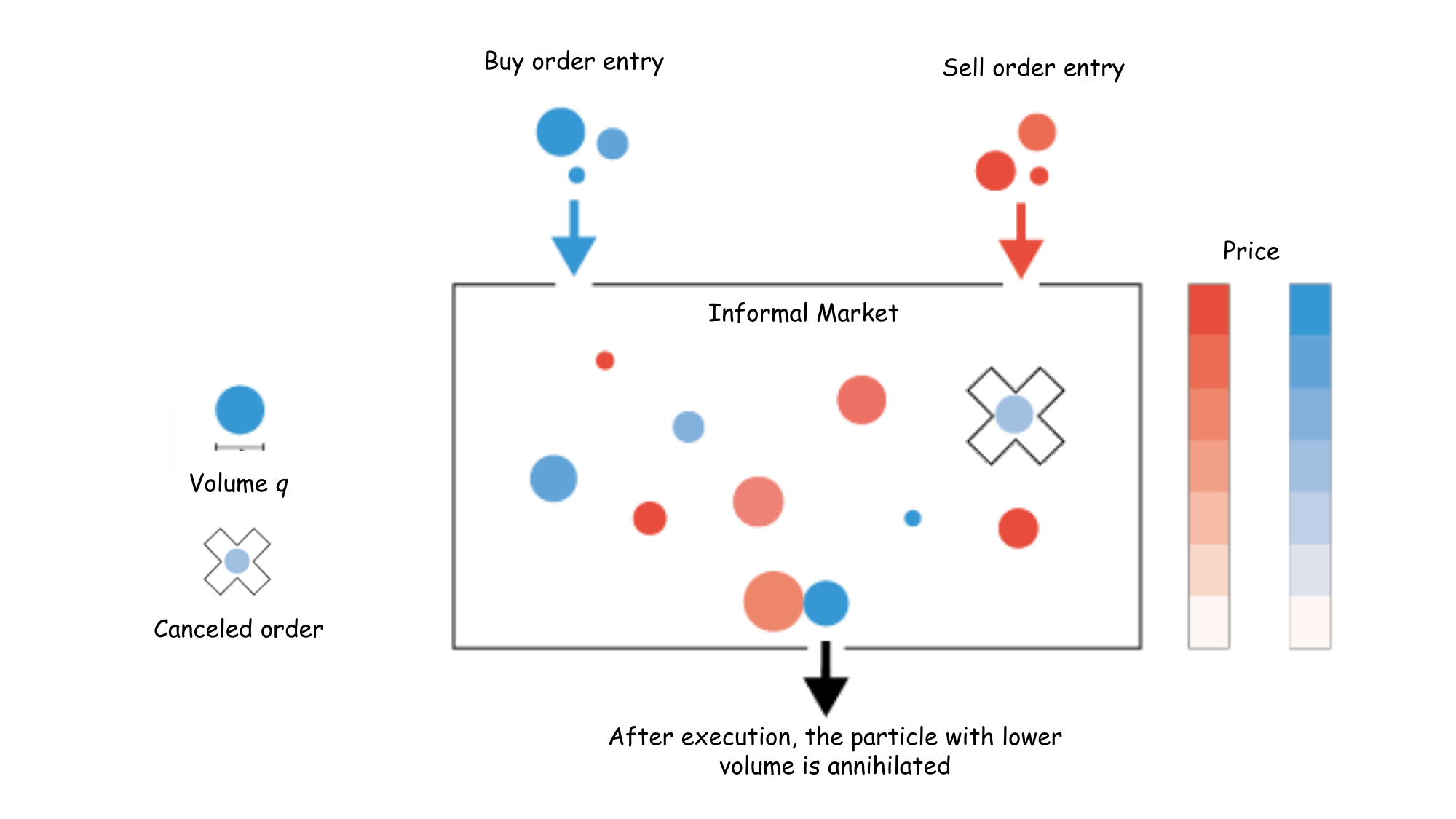}
			\caption{Illustration of the informal market dynamics as a gas of agents with buy or sell intentions. Three processes occur: entry, cancellation, and execution, according to the prices and volumes of the orders.}
			\label{fig:flow chart mercado informal}
		\end{figure}

		The overall dynamics are schematized in Figure \ref{fig:flow chart mercado informal}, where the market is represented as a gas of agents that enter, cancel, or interact through executions, determined by price and volume attributes.
		
	\subsection{Master Equation}
	
		It is possible to formulate a master equation to describe the temporal evolution of the order quantity profile $N(p,q,t)$, without distinguishing between buy or sell orders, considering the processes of entry, execution, and cancellation, in that order:
		
		\begin{equation} \label{ec: Master Equation original mercado informal resumida}
			\frac{d}{dt} N(p, q, t) = \phi P(p, q, t) + M(p, q, t) - \Delta C(p, q, t)
		\end{equation}
		
		The term $M(p,q,t)$ includes both the migration of orders to smaller volumes after partial executions and the removal of orders that are matched with compatible counterparts. When distinguishing between buy and sell orders, these processes are expressed as (we remove time dependence on all functions for brevity):
		
		\begin{equation}
		\small
			\label{ec: Expresion para matcheos mercado informal}
			\begin{split}
				M_a(p^*, q^*) =
				& \kappa_a \Bigg[ \int_{p^*}^{\infty} \int_{q^*}^{+\infty} N_a(p^*, q) N_b(p', q - q^*) \, \ud q \, \ud p' \\
				& - N_a(p^*, q^*) \int_{p^*}^{\infty} \int_{0}^{+\infty} N_b(p', q') \, \ud q' \, \ud p' \Bigg]\\
			\end{split}
\end{equation}
\begin{equation}\small
			\begin{split}
				M_b(p^*, q^*) =
				& \kappa_b \Bigg[ \int_{0}^{p^*} \int_{q^*}^{+\infty} N_b(p^*, q) N_a(p', q - q^*) \, \ud q \, \ud p' \\
				& - N_b(p^*, q^*) \int_{0}^{p^*} \int_{0}^{+\infty} N_a(p', q') \, \ud q' \, \ud p' \Bigg]
			\end{split}
		\end{equation}
		
		To reduce complexity, the factorization $P(s(t),p,q) = P(s(t),p)Q(q)$ is introduced, with exponential distributions for the volumes, justified based on empirical evidence from the study of the Cuban informal foreign exchange market, as discussed in \cite{delaolivacuelloModeloGasPara2024}. Thus, the profiles $N_a(p,q,t)$ and $N_b(p,q,t)$ take the form:
		
		\begin{equation}
			\label{ec: N_a(p) y N_b(p) factorizadas mercado informal}
			\begin{aligned}
				N_a(p, q, t) &= Q_a(q) n_a(p, t) = n_a(p, t) \lambda_a e^{-\lambda_a q}\\
				N_b(p, q, t) &= Q_b(q) n_b(p, t) = n_b(p, t) \lambda_b e^{-\lambda_b q}
			\end{aligned}
		\end{equation}
		
		The resulting interaction terms are expressed as:
		
		\begin{equation}
			\label{ec: termino de mathceos factorizado mercado informal}
			\begin{gathered}
				m_a(p,t) = \kappa_a N_b(t) \left(\gamma_a - 1\right) n_a(p,t) \frac{\int_{p}^{\infty} n_b(p',t) \, \ud p'}{N_b(t)}  \\
				m_b(p,t) = \kappa_b N_a(t) \left(\gamma_b - 1\right) n_b(p,t) \frac{\int_{0}^{p} n_a(p',t) \, \ud p'}{N_a(t)}
			\end{gathered}
		\end{equation}
		with
		\[
		\gamma_a = \frac{1}{1 + \frac{\lambda_a}{\lambda_b}} \quad; \quad \gamma_b = \frac{1}{1 + \frac{\lambda_b}{\lambda_a}}
		\]
		
		Finally, the master equation for the dynamics of the order quantities at price $p$ and time $t$ can be summarized as:
		\begin{widetext}
		\begin{equation}
			\label{ec: Master Equation dinamica general mercado informal}
			\begin{aligned}
				\frac{\ud n_a(p,t)}{\ud t} &= 
				\phi_a P_a(s(t), p) - \kappa_a (1 - \gamma_a) n_a(p,t) \int_{p}^{\infty} n_b(p',t) \, dp'
				- \Delta_a n_a(p,t) \\
				\frac{\ud n_b(p,t)}{\ud t} &= 
				\phi_b P_b(s(t), p)
				- \kappa_b (1 - \gamma_b) n_b(p,t) \int_{0}^{p} n_a(p',t) \, dp'
				- \Delta_b n_b(p,t)
			\end{aligned}
		\end{equation}
\end{widetext}

	\subsection{Analytical Solution}
	
		Let us now consider the stationary case, in which the system reaches a regime where the distributions do not vary in time. By setting the temporal derivatives in \eqref{ec: Master Equation dinamica general mercado informal} to zero, we obtain:
		
		\begin{equation}
			\label{ec: caso estacionario mercado informal}
			\begin{aligned}
				\Delta_b n_b(p) &= I_b(p) + \chi_b n_b(p) \int_{0}^{p} n_a(p') \, dp' \\
				\Delta_a n_a(p) &= I_a(p) + \chi_a n_a(p) \int_{p}^{\infty} n_b(p') \, dp'
			\end{aligned}
		\end{equation}
		where \begin{align*}
		I_a(p) = \phi_a P_a(s,p) \qquad & \chi_a = \kappa_a \left(\gamma_a - 1\right) \\
		I_b(p) = \phi_b P_b(s,p)  \qquad & \chi_b = \kappa_b \left(\gamma_b - 1\right).
		      \end{align*}
		Defining $m(p) = n_a(p) n_b(p)$ and after some algebraic manipulations, we arrive at the differential equation:
		\[\frac{dm(p)}{dp} = m^2(p)\bigg[ \frac{\chi_b}{I_b(p)} - \frac{\chi_a}{I_a(p)} \bigg] + m(p) \bigg[ \frac{I_a'(p)}{I_a(p)} + \frac{I_b'(p)}{I_b(p)} \bigg].\]
		
		The solution of this equation is done in appendix \ref{app:solution}, resulting in:
		\begin{equation}
			\begin{aligned}
				n_a(p) &= C_a \sqrt{\frac{m(p)}{C}} 
				\exp\left(-X \int_{0}^{p} \frac{m(p')}{I(p')} \, dp' \right) \\
				n_b(p) &= C_b \sqrt{\frac{m(p)}{C}} 
				\exp\left(X \int_{0}^{p} \frac{m(p')}{I(p')} \, dp' \right).
			\end{aligned}
			\label{ec: eq n_a_n_b_with_m}
		\end{equation}
		In the case of Gaussian price distributions, we have:
		\begin{equation}
			\label{eq:n_a_n_b_soluciones_compactas}
			\small
			\begin{aligned}
				n_a(p) &= C_a \sqrt{ 
					\exp\left( -\frac{(p' - \mu)^2}{\sigma^2} \Big|_{0}^{p} \right)
				} \\
				&\quad \times \exp\left(
				-C X \int_{0}^{p} 
				\frac{
					\exp\left( -\frac{(p'' - \mu)^2}{\sigma^2} \Big|_{0}^{p'} \right)
				}{
					\tfrac{1}{\sqrt{2\pi \sigma^2}} 
					\exp\left( -\tfrac{(p' - \mu)^2}{2 \sigma^2} \right)
				}
				\ud p'
				\right) 
				\\[1em]
				n_b(p) &= C_b \sqrt{ 
					\exp\left( -\frac{(p' - \mu)^2}{\sigma^2} \Big|_{0}^{p} \right)
				} \\
				&\quad \times \exp\left(
				C X \int_{0}^{p} 
				\frac{
					\exp\left( -\frac{(p'' - \mu)^2}{\sigma^2} \Big|_{0}^{p'} \right)
				}{
					\tfrac{1}{\sqrt{2\pi \sigma^2}} 
					\exp\left( -\tfrac{(p' - \mu)^2}{2 \sigma^2} \right)
				}
				\ud p'
				\right)
			\end{aligned}
		\end{equation}
		
	\subsection{Simulation and Comparison}
	
		To validate the model, simulations were carried out using two complementary approaches based on Monte Carlo techniques, restricted to the case with symmetric parameters and Gaussian-distributed prices.
		
		First, an ABM was developed in \textit{Python}, where buy and sell orders enter at rate $\phi$ and are canceled with probability $\Delta$, as shown in Algorithm \ref{alg:LOB} with the only difference being the removal of restrictions during the initial sampling phase. Prices were sampled from a normal distribution $P(s(t),\sigma^2,p)$ centered at $s(t)$ with unit variance, while volumes were fixed at $q=1$ for simplicity. The price space was discretized over the interval $s(t) \pm 5\sigma$, divided into 500 \textit{bins}. At each iteration, an agent interacts with a fraction $\kappa$ of agents of the opposite type, executing transactions when $p_b \geq p_a$. To simplify the dynamics, $s=100$ was held constant over time.
		
		In parallel, a stochastic reaction model was implemented using the Gillespie algorithm \cite{gillespieExactStochasticSimulation2002}, with the \textit{StochPy} package \cite{maarleveldStochPyComprehensiveUserFriendly2013}, as shown in Algorithm \ref{alg:gillespie} where for this scenario $\psi = 0$. The price space $[100 \pm 5\sigma]$ was discretized into 300 \textit{bins}, interpreted as species participating in three possible reactions: entry, cancellation, and interaction. The associated rates were $\phi P(100,\sigma^2,p)$ (entry), $\Delta n(p,t)$ (cancellation), and $\kappa n(p,t) n_\text{opposite}(p,t)$ (interaction), analogous to those in the ABM.
		
		Additionally, the master equation \eqref{ec: Master Equation dinamica general mercado informal} was numerically integrated over $T=1000$ with step $dt=0.1$, using the Runge–Kutta 5(4) method implemented in the \textit{SciPy} package. A discretization of 1000 \textit{bins} was used, with parameters $\phi=15$, $\kappa=0.01$, and $\Delta=0.01$, selected following criteria similar to those presented in section \ref{sub:Simulacion LOB}.
		
		Simulations were run for $T=1000$, starting from Gaussian distributions centered at $s=100$. To extract the stationary profiles $n_a(p)$ and $n_b(p)$, the last 750 histograms after reaching equilibrium were averaged. Ten simulations were performed with the ABM and five with the Gillespie algorithm, with results averaged accordingly. In the latter case, trajectories were re-scaled to a common time frame for averaging.
		
		\begin{figure}
			\centering
			\includegraphics[width=\columnwidth, keepaspectratio]{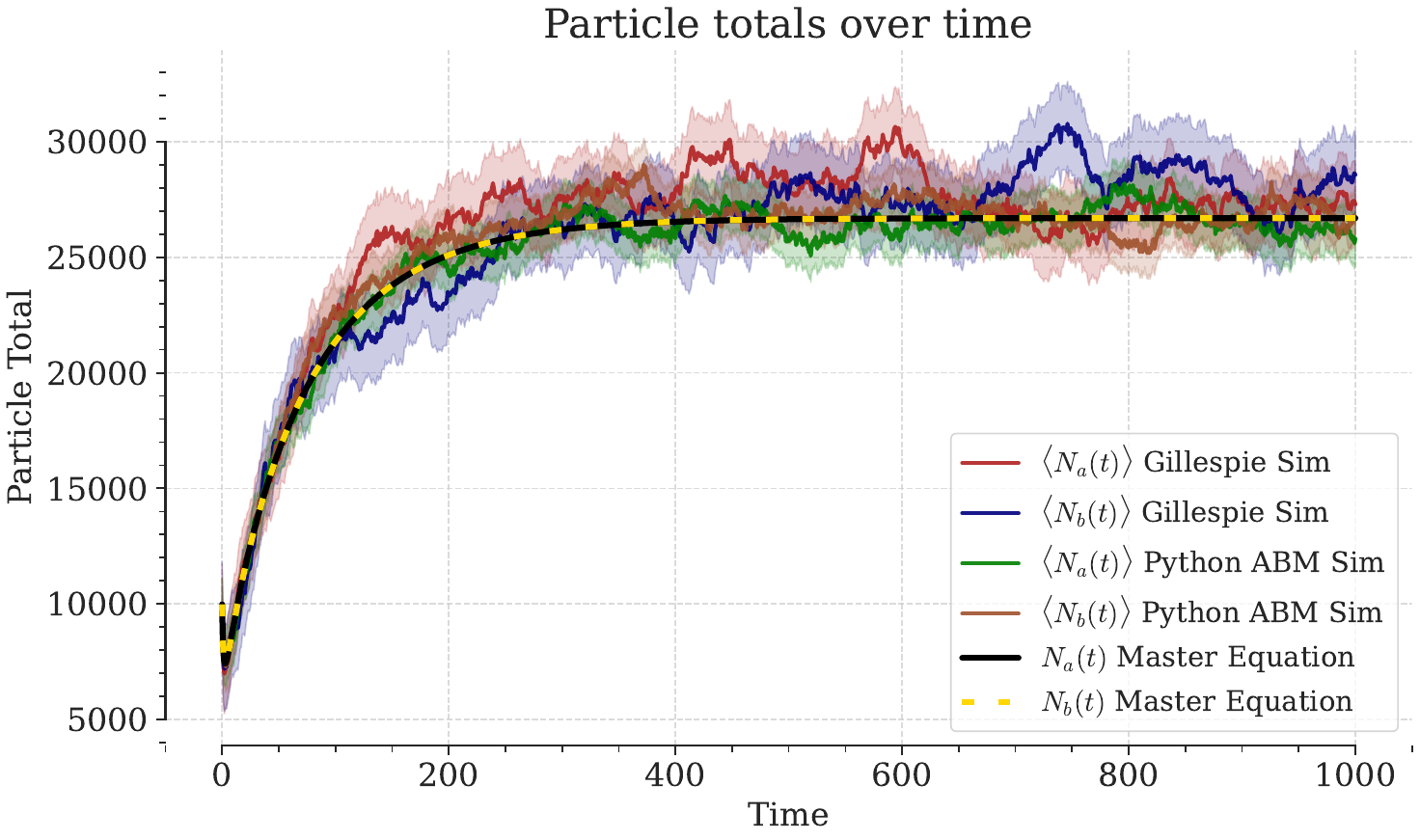}
			\caption{Average total buy and sell orders as a function of time, obtained through stochastic simulation and numerical integration of the master equation \ref{ec: Master Equation dinamica general mercado informal}. Error bands corresponding to one standard deviation are also shown.}
			\label{fig:totales de particulas mercado informal}
		\end{figure}
		
		\begin{figure*}
			\centering
			\includegraphics[width=0.8\textwidth, keepaspectratio]{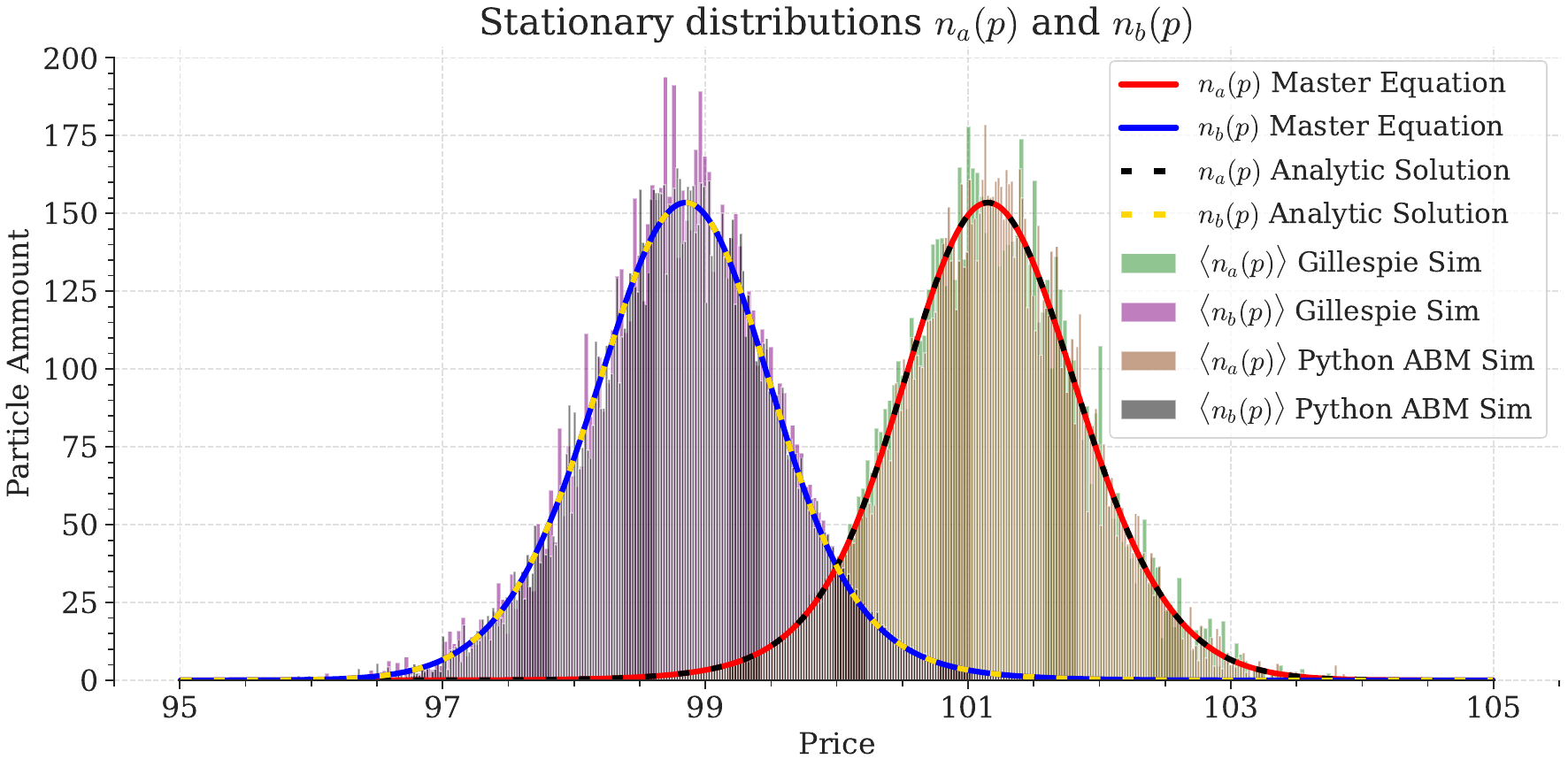}
			\caption{Averaged stationary profiles of $n_a(p)$ and $n_b(p)$ obtained via stochastic simulation, numerical integration of the master equation \ref{ec: Master Equation dinamica general mercado informal}, and analytical solution for the symmetric Gaussian case \ref{eq:n_a_n_b_soluciones_compactas}.}
			\label{fig:perfiles na y nb mercado informal}
		\end{figure*}
		
		Figures \ref{fig:totales de particulas mercado informal} and \ref{fig:perfiles na y nb mercado informal} show the strong agreement between both methodologies, both in the evolution of total orders and in the stationary profiles. Although the master equation does not capture stochastic fluctuations, it accurately reproduces the average evolution.
		
		Finally, the analytical solutions of \eqref{ec: solucion estacionaria para m(p) mercado informal} and \eqref{ec: soluciones estacionarias para n_a(p) y n_b(p) mercado informal} match the numerical profiles, as shown in Figure \ref{fig:perfiles na y nb mercado informal}. For the integration constants, we obtained $C \approx 1.94\times 10^{-8}$, $C_a \approx 0.85\times 10^{-5}$, and $C_b \approx 2.29\times 10^{-3}$, confirming the relation $C \approx C_a C_b$ within the margin of error.
		
\section{Formal-informal connection}
	
	Up to this point, we have developed two similar methodologies that differ in their essence, to characterize the dynamics of both types of markets, formal and informal, using mean-field treatments based on master equations. In this section, we demonstrate that the equations of the informal market can be adapted to also describe the organized dynamics of the formal market in an appropriate limit.
	
	\subsection{Physical Interpretation}
	
		In the stationary regime, the gas of agents describing the informal market can be understood as an open system, composed of a classical gas with two types of particles: buyers and sellers. These occupy discrete price levels without internal movement between them, so that price does not act as a dynamic degree of freedom. Particles leave the system only via cancellation or execution.
			
		The system operates at fixed temperature $T$ and volume $V$, although both are secondary for the dynamics. The temperature, expressed as $\beta \sim 1/T$, can be thought of as related to the amplitude of fluctuations, while the constant volume implies that higher gas density leads to more interactions. With these considerations, the use of a grand–canonical treatment is justified, even in the absence of explicit dynamics.
			
		An effective interaction energy $E_{eff}$ is introduced, associated with the cost of coexistence of opposite-type particles under execution conditions, which drives energy minimization through the exhaustion of these possible interactions. Additionally, effective chemical potentials $\mu_{eff_A}(p)$ and $\mu_{eff_B}(p)$ are defined for sell and buy orders. These set the relationship between entry and cancellation rates at each price level, being “effective” in the sense that their formulation differs from traditional Statistical Physics. Each price level has its own effective potential, dependent on the local particle number.
		
		The grand–canonical distribution $P[n_a,n_b]$ of the stationary state takes the form:
		
		\begin{align}\small
			P[n_a,n_b]
			&= \frac{1}{\Xi} 
			\exp\Bigg\{-\beta \Bigg[E_{eff} \\
			&- \int_{0}^{\infty} \mu_{eff_A}(p)\, n_a(p)\, dp \notag\\
			&- \int_{0}^{\infty} \mu_{eff_B}(p)\, n_b(p)\, dp \Bigg]\Bigg\}
			\label{ec: gran canonica informal}
		\end{align}
		
		From the stationary equations \eqref{ec: caso estacionario mercado informal}, an effective functional $L[n_a,n_b]$ can be defined, analogous to a system Lagrangian, whose minimization reproduces the stationary conditions:
		
		\begin{equation} \label{ec:accion efectiva sistema informal}
			\begin{split}
				L[n_a, n_b]  
				&= \frac{1}{1-\gamma_a} \int_{0}^{\infty} \left[ \phi_a P_a(s,p) \ln n_a(p) - \Delta_a n_a(p) \right] dp \\
				& + \frac{1}{1-\gamma_b} \int_{0}^{\infty} \left[ \phi_b P_b(s,p) \ln n_b(p) - \Delta_b n_b(p) \right] dp \\
				& - \kappa \int_{0}^{\infty} dp_a \int_{0}^{\infty} \Theta(p_b - p_a) n_a(p_a) n_b(p_b) \, dp_b
			\end{split}
		\end{equation}
		
		From this, some effective thermodynamic quantities can be identified, such as an interaction energy:
		
		\begin{equation}
			E_{eff} = \kappa \int_{0}^{\infty} dp_a \int_{0}^{\infty} \Theta[p_b - p_a] n_a(p_a) n_b(p_b) dp_b
			\label{ec: Hamiltoniano informal}
		\end{equation}
		
		and the effective chemical potentials:
		
		\begin{equation}
			\small
			\begin{aligned}
				\int_{0}^{\infty} \mu_{eff_A}(p) n_a(p) dp
				&= \frac{1}{1-\gamma_a} \int_{0}^{\infty} \left[\frac{\phi_a P_a(s,p) \ln n_a(p)}{ n_a(p)}  - \Delta_a \right]\\
				&\quad \times n_a(p) dp \\
				\int_{0}^{\infty} \mu_{eff_B}(p) n_b(p) dp
				&= \frac{1}{1-\gamma_b} \int_{0}^{\infty} \left[\frac{\phi_b P_b(s,p) \ln n_b(p)}{n_b(p)}  - \Delta_b \right]\\
				&\quad \times n_b(p) dp
			\end{aligned}
			\label{ec: potenciales_quim}
		\end{equation}
		
		By analogy with the standard treatment in Statistical Physics for such systems, the particle entry and cancellation terms have been separated within the effective chemical potential, and the interaction term between particles is defined separately as an effective energy.
		\begin{figure}[!htb]
			\centering
			\includegraphics[width=\columnwidth]{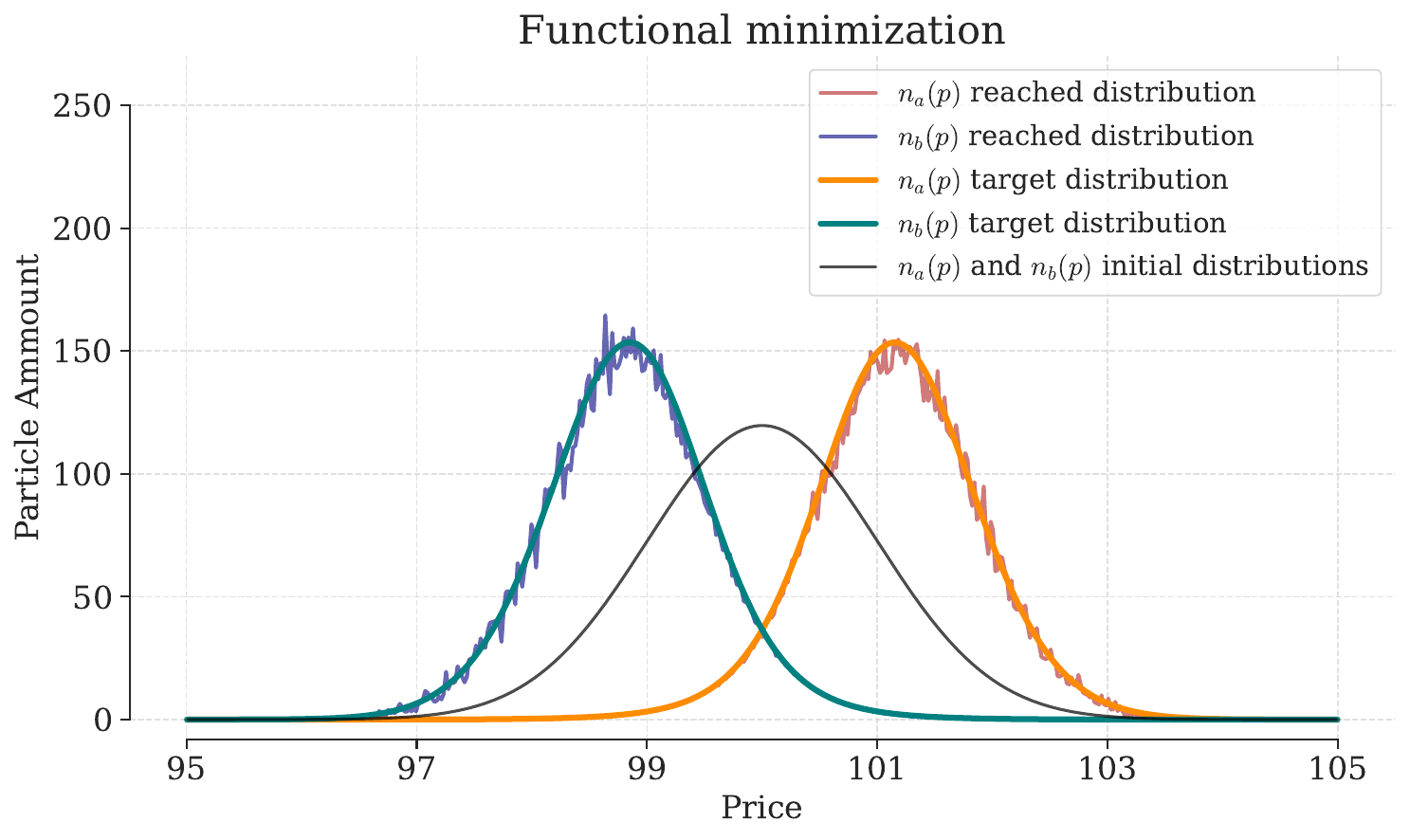}
			\caption{Verification using the Metropolis Algorithm of the convergence of the informal system to the actual equilibrium distributions. The symmetric Gaussian case is employed.}
			\label{fig:Metropolis mercado informal}
		\end{figure}

		To validate this approach, the Metropolis Algorithm was implemented in \textit{Python} to minimize the functional $L[n_a,n_b]$, using $\phi=10$, $\kappa=0.05$, and $\Delta=0.01$ in the symmetric case with Gaussian prices centered at $s=100$ and unit variance. As shown in Figure \ref{fig:Metropolis mercado informal}, the results converge to the equilibrium distributions, confirming the validity of the proposed grand–canonical treatment.

		\subsection{Generalized Master Equation}
	
		The expressions for the effective chemical potentials obtained for the informal system \eqref{ec: potenciales_quim} can be extended to the LOB case, since both systems share order entry dynamics following Poisson processes and cancellation at constant rates. The essential difference lies in the energetic term that regulates interactions.
		
		In a LOB-type system, executions do not occur randomly; rather, orders with higher mutual attractiveness, meaning prices closer to the optimal execution, are prioritized. To capture this selectivity, a preferential interaction parameter $\psi$ is introduced, which modifies the energetic cost of interaction according to the price difference $p_b - p_a$. This factor is incorporated through an exponential weight $e^{\psi (p_b - p_a)}$, increasing the interaction probability as mutual attractiveness grows. The modified effective energy term is expressed as:
		
		\begin{equation}
			E_{eff} = \kappa \int_{0}^{\infty} dp_a \int_{0}^{\infty} \Theta[p_b - p_a] e^{\psi (p_b - p_a)} n_a(p_a) n_b(p_b) dp_b
			\label{ec: Hamiltoniano con psi}
		\end{equation}
		
		The informal market dynamics are recovered for $\psi = 0$, while in the limit $\psi \to \infty$, non-optimal interactions are strongly penalized, and only those with maximal price difference are executed.
		
\begin{figure}[!htb]
			\centering
			\includegraphics[width=\columnwidth]{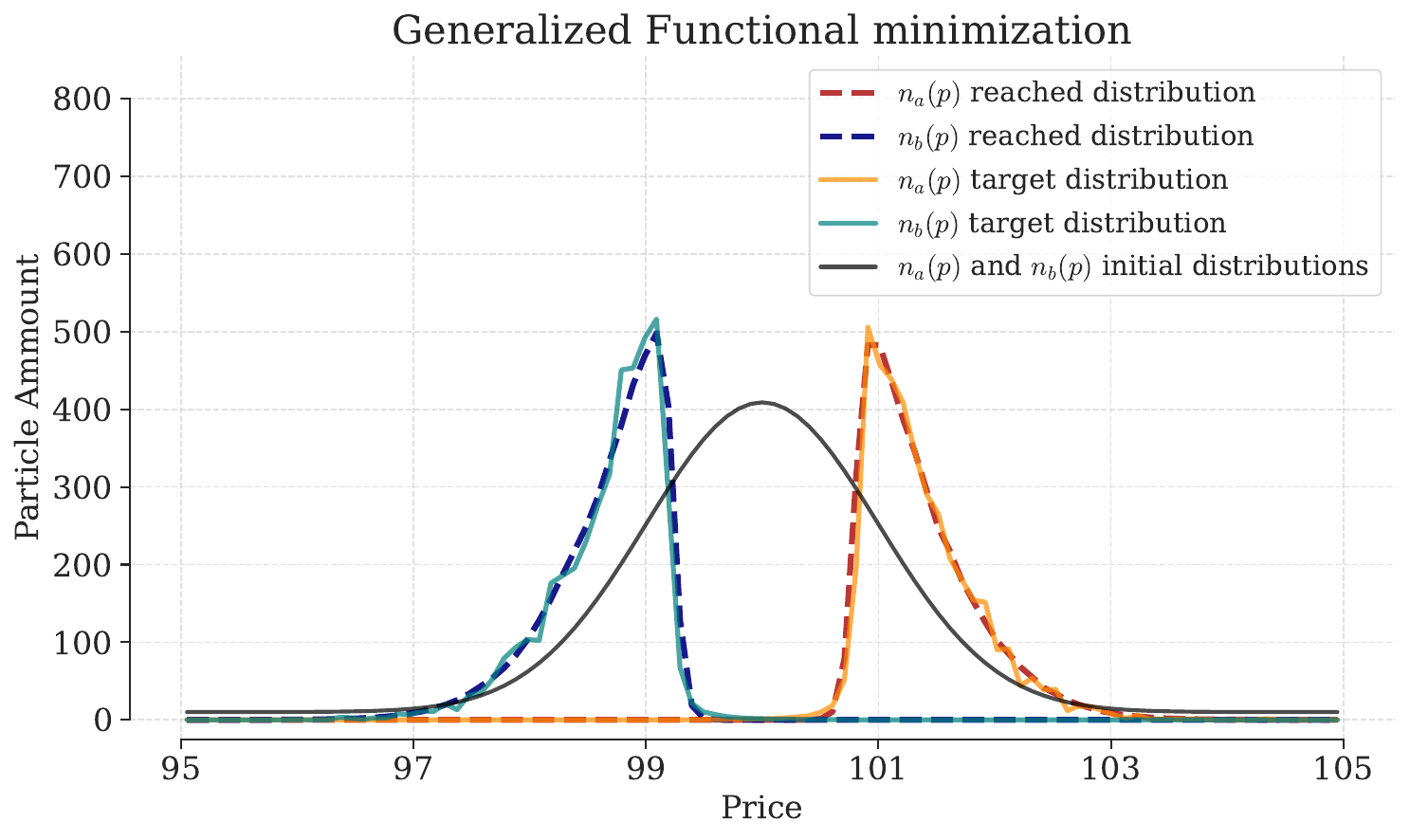}
			\caption{Verification using the Metropolis Algorithm of the convergence of the dynamic system with a high preferential parameter $\psi$ toward the equilibrium distribution associated with the LOB.}
			\label{fig:Metropolis con psi}
		\end{figure}

		This proposal was validated by implementing the Metropolis Algorithm in \textit{Python}, using symmetric parameters favoring optimal executions: $\psi = 20$, $\phi = 20$, $\kappa = 10^{10}$, and $\Delta = 0.01$, with Gaussian prices centered at $s=100$ and unit variance. As shown in Figure \ref{fig:Metropolis con psi}, convergence toward the equilibrium distributions characteristic of the LOB is observed. The choice of a high $\kappa$ ensures that each possible preferential interaction occurs immediately with high probability.
		
		From this modified energetic term, the master equation of the informal system \eqref{ec: Master Equation dinamica general mercado informal} is generalized to include the exponential factor that redistributes the execution probability in the interaction term. Thus, the temporal evolution of the order quantity profiles $n_a(p,t)$ and $n_b(p,t)$ is described by:
		
		\begin{widetext}
		\begin{equation} \label{ec:ecuacion maestra gas con psi}
			\begin{aligned}
				\frac{\ud n_a(p,t)}{\ud t} &= \phi_a P_a(s(t), p)
				 - \Delta_a n_a(p,t)
				 - \kappa (1 - \gamma_a)\, e^{-\psi p} n_a(p,t) \int_{p}^{\infty} e^{\psi p'} n_b(p',t) \, dp' \\
				\frac{\ud n_b(p,t)}{\ud t} &= \phi_b P_b(s(t), p)
				 - \Delta_b n_b(p,t)
				 - \kappa (1 - \gamma_b)\, e^{\psi p} n_b(p,t) \int_{0}^{p} e^{-\psi p'} n_a(p',t) \, dp'
			\end{aligned}
		\end{equation}
		\end{widetext}

		Each term retains the interpretation given for the informal system, except that executions are modulated by an exponential weight favoring optimal interactions, effectively acting as a larger cross-section for orders with more attractive prices.

	\subsection{Simulation and Comparison}
	
		To validate the generalized model with the preferential parameter $\psi$, simulations were carried out using both an agent-based model (ABM) implemented in \textit{Python} (Algorithm \ref{alg:LOB}) and a stochastic simulation based on the Gillespie Algorithm via \textit{StochPy} (Algorithm \ref{alg:gillespie}). In both approaches, symmetric parameters were used: $\phi = 20$, $\Delta = 0.01$, $\psi = 20$, and $\kappa = 10^{10}$, ensuring the immediate execution of all optimal interactions. The values of $\psi$ and $\kappa$ were adjusted considering computational constraints and the numerical stability of the exponential terms in the generalized equations.
		\begin{figure*}[!htb]
			\centering
			\includegraphics[width=0.8\textwidth]{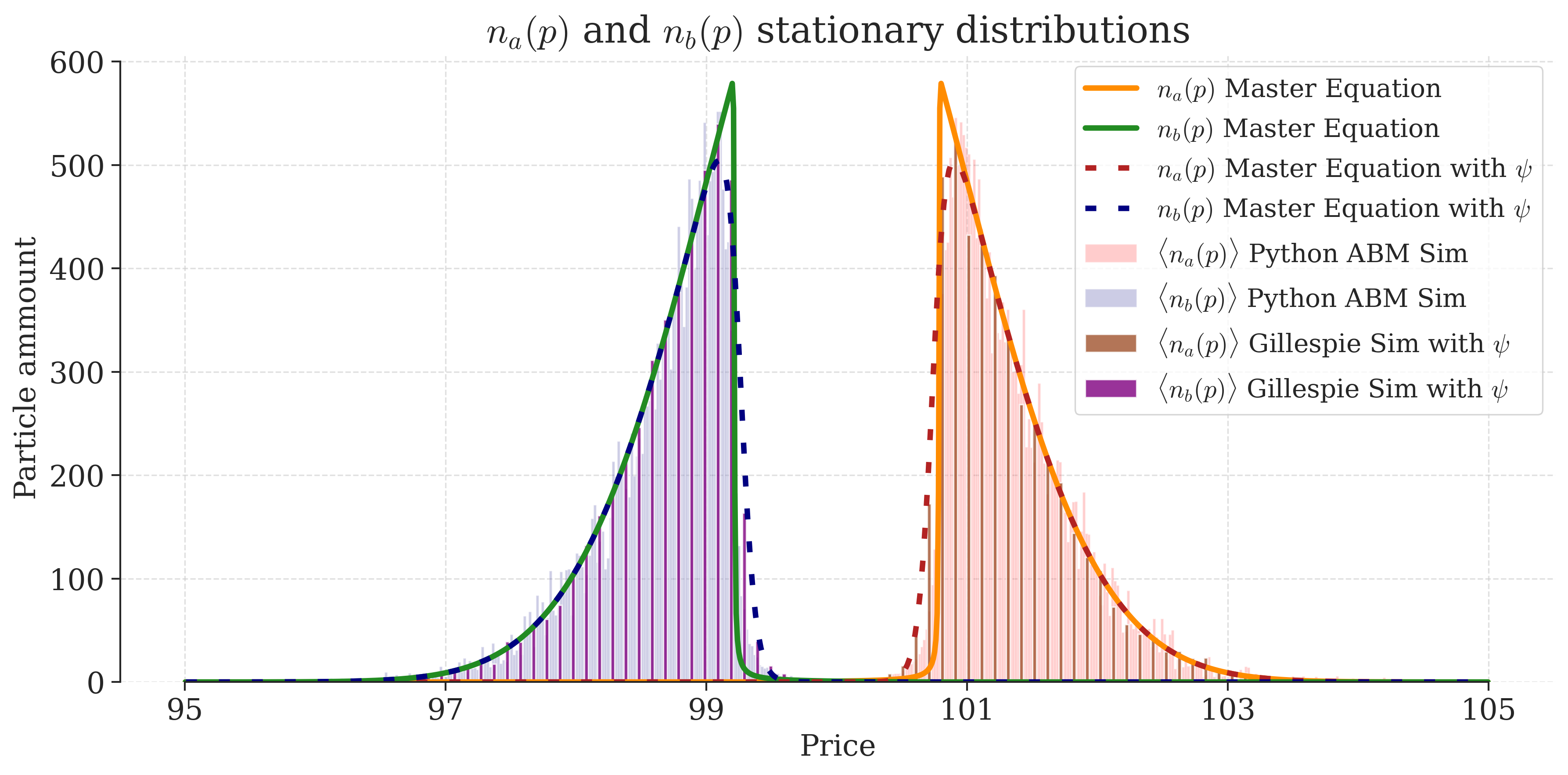}
			\caption{Average stationary profiles of $n_a(p)$ and $n_b(p)$ obtained from the LOB simulation and from the numerical integration of the system’s master equation \ref{ec:ecuacion maestra gas con psi} with $\psi = 20$ for the symmetric Gaussian case.}
			\label{fig:perfiles estacionarios LOB con psi}
		\end{figure*}

		A symmetric Gaussian price distribution centered at $s = 100$ with variance $\sigma^2 = 1$ was used. For the Gillespie implementation, the discrete grid included 100 \textit{bins} within $[s - 5\sigma, s + 5\sigma]$, while integration of the master equation \eqref{ec:ecuacion maestra gas con psi} was performed using the Radau method, increasing the resolution to 500 \textit{bins} to improve precision and ensure convergence. The time step was $dt = 0.1$, and the total duration $T = 1500$, averaging the last 750 histograms per simulation to estimate the stationary profiles $n_a(p)$ and $n_b(p)$.

		\begin{figure}
			\centering
			\includegraphics[width=\columnwidth]{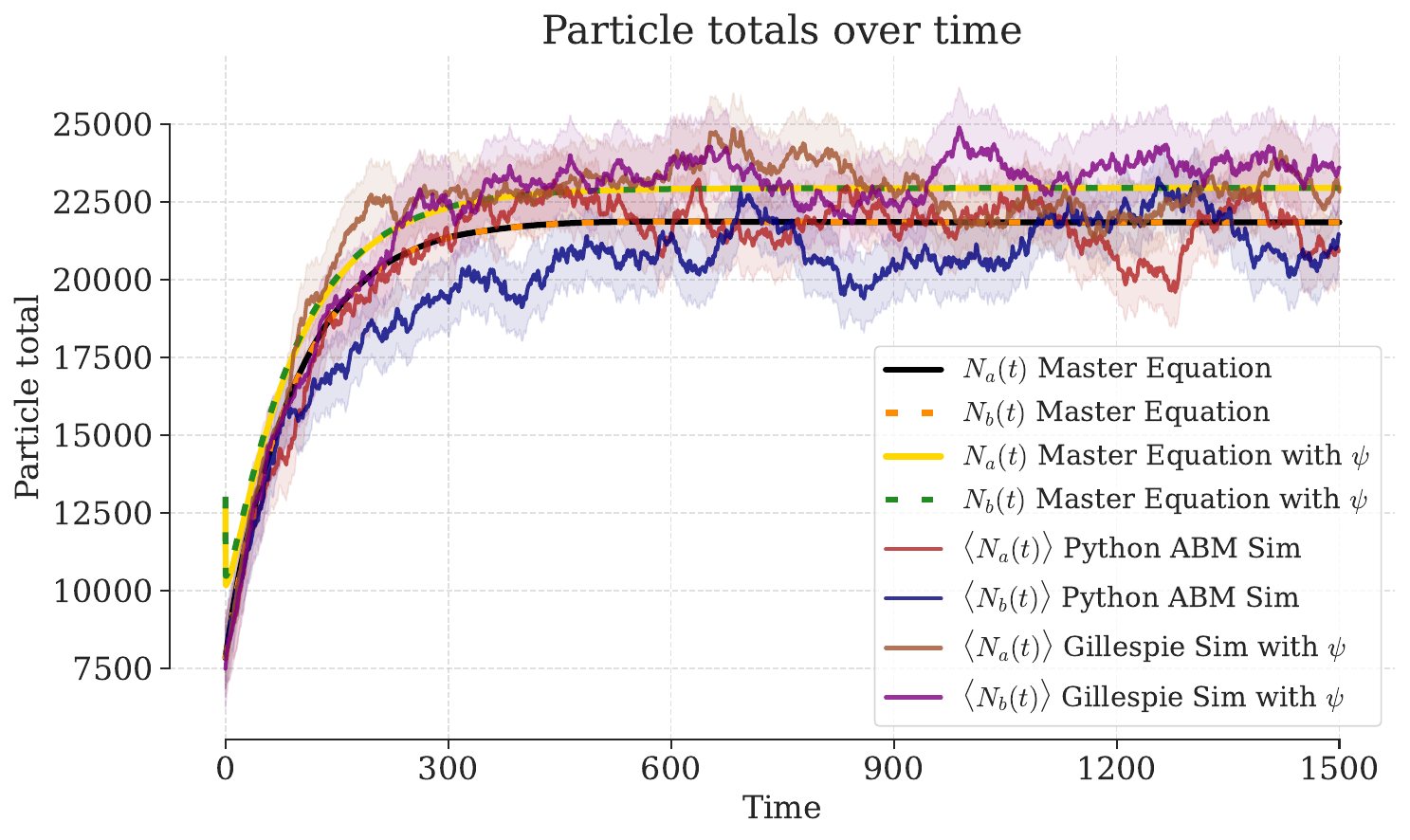}
			\caption{Average total orders obtained from the LOB simulation and from the numerical integration of the system’s master equation \ref{ec:ecuacion maestra gas con psi} with $\psi = 20$ for the symmetric Gaussian case.}
			\label{fig:totales LOB con psi}
		\end{figure}

		The results, shown in Figures \ref{fig:totales LOB con psi} and \ref{fig:perfiles estacionarios LOB con psi}, reveal excellent agreement between both methodologies, both in the temporal evolution of the total number of orders and in the stationary profiles. A slight overestimation of the total number of orders is observed relative to the predictions obtained from the probabilistic model of the LOB, attributed to the parametric sensitivity of the generalized model with respect to $\psi$ and $\kappa$.
		
		Finally, it is worth mentioning that the accuracy of the results is expected to improve when using finer discretizations along the price axis, since the discrete representation introduces quantization errors when approximating the differentials $dp$ from the continuous formulation, affecting both the simulation and the numerical integration.

	\subsection{Effective Temperature}
	
		The equations \eqref{ec: Master Equation dinamica general mercado informal}, being a mean-field approach independent of the total number of orders, are insensitive to system fluctuations. This approximation is suitable in simulations with a large number of particles, where relative fluctuations are negligible.
		
		However, for a hypothetical LOB with a low number of orders, the generalized master equation \eqref{ec:ecuacion maestra gas con psi} with a finite and small preferential parameter $\psi$ allows capturing the fluctuating effects of the system, with $\psi$ interpreted as an effective temperature. Although a systematic analysis of the equivalence between this effective temperature and the system size was not performed in this research, we want to illustrate its feasibility by fitting the best value of $\psi$ for a small system.  

		\begin{figure}
			\centering
			\includegraphics[width=\columnwidth]{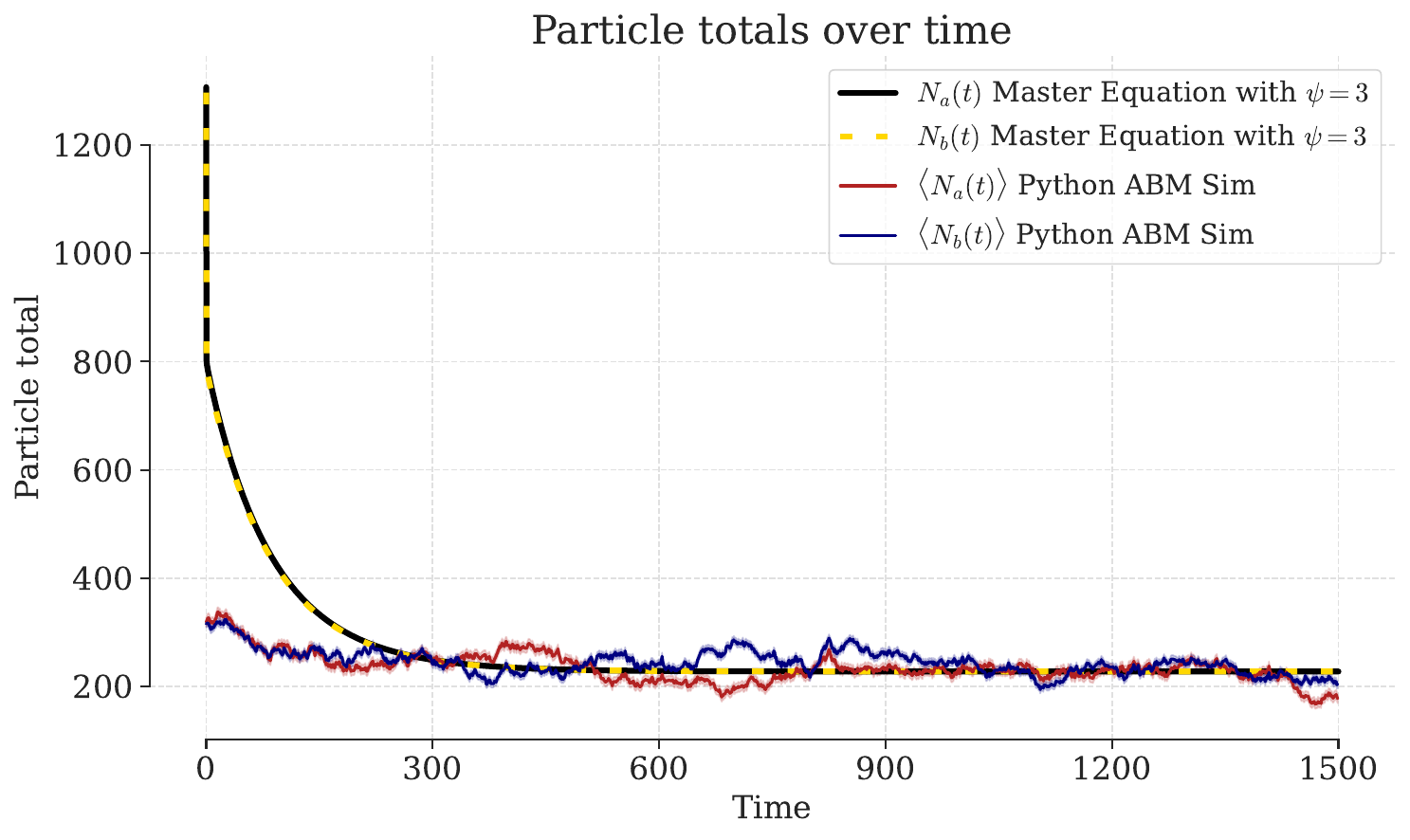}
			\caption{Total orders obtained during the LOB simulation with few particles, along with those resulting from the numerical integration of the system’s master equation \ref{ec:ecuacion maestra gas con psi} with $\psi = 3$ for the symmetric Gaussian case.}
			\label{fig:totales LOB pocas particulas con psi}
		\end{figure}

		\begin{figure}
			\centering
			\includegraphics[width=0.5\textwidth]{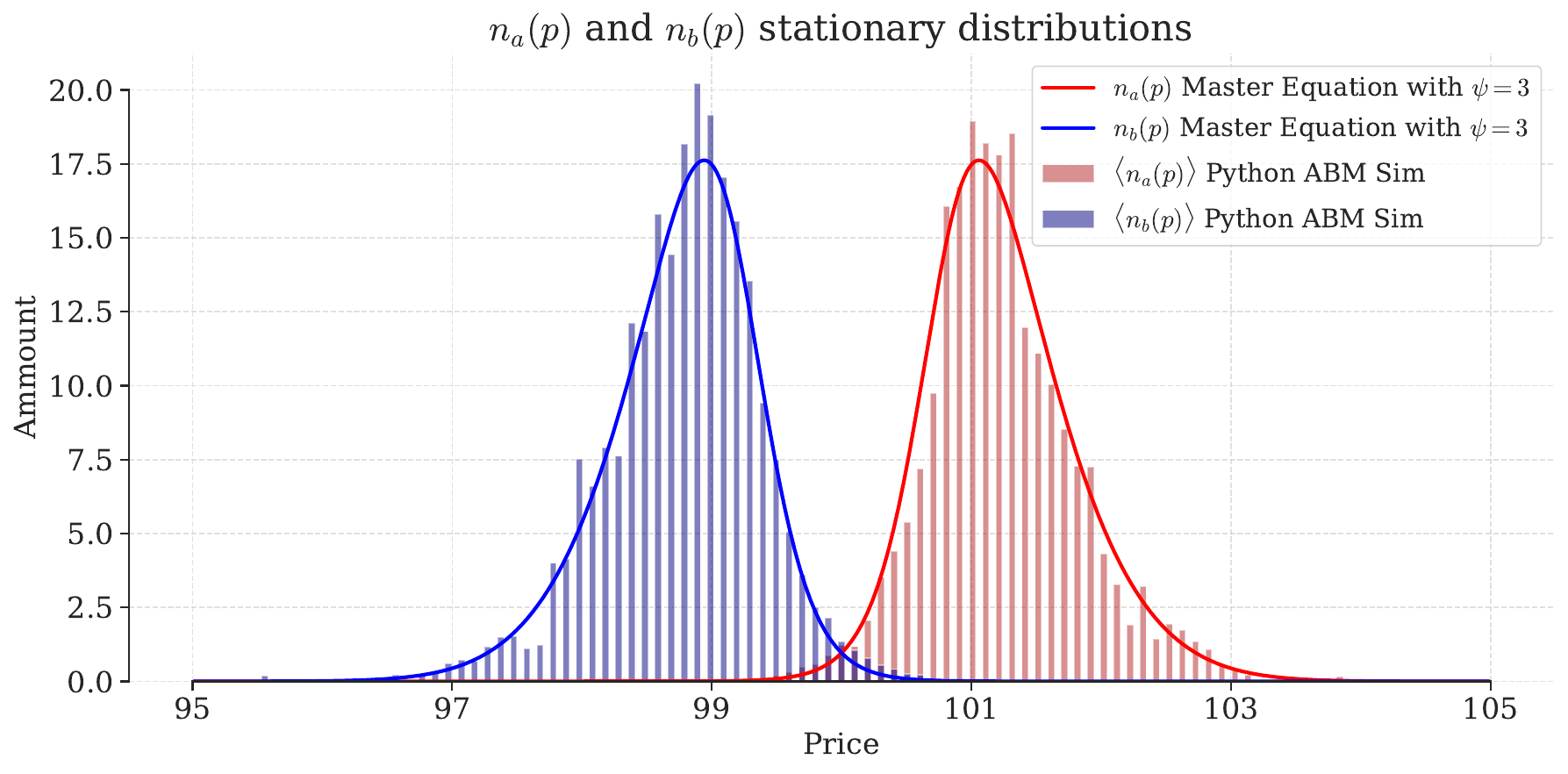}
			\caption{Stationary profiles of $n_a(p)$ and $n_b(p)$ obtained from the LOB simulation with few particles, along with those resulting from the numerical integration of the system’s master equation \ref{ec:ecuacion maestra gas con psi} with $\psi = 3$ for the symmetric Gaussian case.}
			\label{fig:perfiles estacionarios LOB pocas particulas con psi}
		\end{figure}

		Although a systematic analysis of the equivalence between this effective temperature and the system size is not done at this point, we want to illustrate its feasibility by fitting the best value of $\psi$ for a small system. Figures \ref{fig:totales LOB pocas particulas con psi} and \ref{fig:perfiles estacionarios LOB pocas particulas con psi} show the results for a LOB with few particles under symmetric Gaussian conditions, using $\phi = 1$, $\Delta = 0.01$, $\psi = 3$, and $\kappa = 0.7$, with prices centered at $s = 100$ and variance $\sigma^2 = 1$.

\section{Conclusions}

This work establishes that the dynamics of two idealized market archetypes—fully informal (e.g., fragmented black markets) and fully formal (e.g., Limit Order Book exchanges)—can be described within mean-field frameworks governed by master equations. The key insight is that these regimes are not structurally disjoint but lie at opposite ends of a continuous spectrum, parametrized by a preferential interaction strength $\Psi$. When $\Psi=0$, agents interact indiscriminately, yielding the uncoordinated matching characteristic of informal markets; as $\Psi$ increases, interactions become increasingly selective, favoring price-compatible trades and converging toward LOB-like dynamics in the $\Psi\to \infty$ limit.

Crucially, the master equation for the informal limit reproduces empirical regularities observed in real-world informal markets—most notably, the exponential distribution of transaction-intention sizes documented in the Cuban foreign exchange black market. This agreement suggests that the proposed gas-like formalism captures essential mechanisms of decentralized exchange under minimal institutional structure, offering a physically grounded null model for informal price formation.

More broadly, the framework implies that real markets need not be classified as purely formal or informal, specially nowadays when even informal markets resort to digital platforms. Therefore, having a continuous parameter that interpolates between the two extreme ideal cases can be useful. Informal markets may operate at intermediate values of $\Psi$, reflecting varying degrees of information transparency, price discovery efficiency, or coordination infrastructure. This opens the possibility of quantifying the ``formality'' of any market via effective estimation of $\Psi$ from transaction or order data.

Nevertheless, significant theoretical questions remain open. First, although simulations confirm that the $\Psi\to\infty$ limit of the generalized master equation reproduces LOB-like stationary profiles, we have not succeeded in analytically reducing this limit to the master equation independently derived for the formal LOB (Section II). The two equations share the same phenomenology but differ in functional form as a consequence of inherently different derivations. Resolving this discrepancy would be nice to claim a true unification.

Second, the current approach is strictly mean-field and thus neglects fluctuations arising from finite population size, discreteness, and intrinsic stochasticity. While we observe that a finite $\Psi$ can phenomenologically mimic some features of small-NN LOB simulations—hinting at an interpretation of $\Psi$ as an effective inverse temperature—a systematic mapping between system size, fluctuation magnitude, and optimal $\Psi$ has not been established. A fluctuation-corrected theory (e.g., via system-size expansion or Langevin approximations) would clarify whether $\Psi$ genuinely encodes thermodynamic temperature or merely serves as a fitting parameter.

Finally, real LOBs exhibit power-law distributed order sizes, inter-trade times, and price impacts—statistical signatures that are absent in our Gaussian/mean-field setup. It remains an open question whether the current framework can be extended to generate such heavy-tailed behavior. If not, the model may be fundamentally limited to describing only the ``core'' of the order book, away from the tails where rare but consequential events occur.

Future work should face these challenges, test the model against empirical data across market contexts and push the formalism beyond mean-field to confront the statistical complexity of real financial exchanges.

	\vspace{2em}
	
	\begin{acknowledgments}
		To Prof. Rajiv Sethi, whose conversations with Alejandro Lage Castellanos during the \textit{Complexity Global School} (CGS-24) inspired several of the questions that shaped this research.
		
		Finally, to Alejandro García Figal and Roberto Mulet, for their support and various contributions to this research.
	\end{acknowledgments}
\appendix
\section{Analytical solution of symmetric case}
\label{app:solution}
		Defining the auxiliary functions:
		\[
		f(p) = \frac{\chi_b}{I_b(p)} - \frac{\chi_a}{I_a(p)}\quad , \quad
		g(p) = \frac{I_a'(p)}{I_a(p)} + \frac{I_b'(p)}{I_b(p)}
		\]
		we obtain a Bernoulli equation for $m(p)$:
		\begin{equation} \label{ec: Bernoulli m(p)}
			\frac{dm(p)}{dp} + [-g(p)]m(p) = f(p) m^2(p)
		\end{equation}

		The general solution is given by:
		\begin{equation} \label{ec: solucion estacionaria para m(p) mercado informal}
			m(p) = n_a(p) n_b(p) = \frac{e^{\int_{0}^{p} g(p')dp'}} {\int_{0}^{p} [ \frac{\chi_a}{I_a(p')} - \frac{\chi_b}{I_b(p')}] e^{\int_{0}^{p'} g(p'')dp''}dp' + C}
		\end{equation}

		Finally, the stationary solutions for $n_a(p)$ and $n_b(p)$ are obtained:
		\begin{equation}
			\label{ec: soluciones estacionarias para n_a(p) y n_b(p) mercado informal}
			\begin{aligned}
				n_a(p) &= C_a \exp\left(
				\int_{0}^{p} \frac{1}{I_a(p')} \left[ I_a'(p') - \chi_a m(p') \right] \ud p'
				\right) \\
				n_b(p) &= C_b \exp\left(
				\int_{0}^{p} \frac{1}{I_b(p')} \left[ I_b'(p') + \chi_b m(p') \right] \ud p'
				\right)
			\end{aligned}
		\end{equation}
		Note that $C$, $C_a$, and $C_b$ are integration constants, which must satisfy the relation $C = C_a C_b$.

		Assuming symmetric parameters for both types of orders, $\chi = \chi_a = \chi_b$ and $I(p) = I_a(p) = I_b(p)$, equation \eqref{ec: Bernoulli m(p)} simplifies to:
		\[	\frac{dm(p)}{dp} = m(p) \bigg[ \frac{I_a'(p)}{I_a(p)} + \frac{I_b'(p)}{I_b(p)} \bigg]
		\]
		which allows direct integration for $m(p)$:
		\begin{equation} \label{ec: ec_caso_part}
			m(p) = n_a(p) n_b(p) =  Ce^{2 \int_{0}^{p} \frac{I'(p')}{I(p')} dp'}
		\end{equation}

\section{Simulation algorithms }

\subsection{Limit order book}

The following pseudocode describes the algorithm used to implement the order book simulation.

		\begin{algorithm}[H]
			\small
			\setstretch{0.9}
			\caption{Stochastic simulation of the Limit Order Book}
			\label{alg:LOB}
			\begin{algorithmic}[1]
				\State \textbf{Parameters:} $n_{\text{steps}}, n_{\text{histograms}}, \text{ref}, \sigma, \phi, \Delta$
				\Statex \textit{\# Initialization}
				\State Generate $N_A$ prices $p_A \sim \mathcal{N}(\text{ref}, \sigma)$ such that $p_A > \text{ref} + 0.5$
				\State Generate $N_B$ prices $p_B \sim \mathcal{N}(\text{ref}, \sigma)$ such that $p_B < \text{ref} - 0.5$
				\State Define $[p_{\min}, p_{\max}]$ and divide it into uniform intervals
				\Statex \textit{\# Simulation}
				\For{$t = 1$ to $n_{\text{steps}}$}
				\State Record total number of particles
				\State Add new particles according to Poisson processes with rate $\phi$
				\For{each new particle $X \in \{A, B\}$}
				\State Draw $p \sim \mathcal{N}(\text{ref}, \sigma)$
				\If{it interacts with a counterpart}
				\State Remove both particles
				\Else
				\State Add particle with price $p$ to side $X$
				\EndIf
				\EndFor
				\State Remove particles with probability $\Delta$
				\If{$t$ belongs to the last $n_{\text{histograms}}$ steps}
				\State Store histogram
				\EndIf
				\EndFor
				\Statex \textit{\# Results}
				\State Compute $n_A(p)$ and $n_B(p)$ as the averages of the last $n_{\text{histograms}}$ histograms
				\State Return $N_A(t)$ and $N_B(t)$ as the total number of particles at each time step $t$
			\end{algorithmic}
		\end{algorithm}

\subsection{Gillespie algorithm for market ABM simulation}

This pseudocode describes the Gillespie simulation of a gas of players with buy/sell offers whose interaction rate favors the interaction of the most attractive prices, with an exponential weight proportional to $\psi (p_j-p_i)$.

		\begin{algorithm}[H]
			\small
			\setstretch{0.9}
			\caption{Stochastic simulation using the Gillespie algorithm}
			\label{alg:gillespie}
			\begin{algorithmic}[1]
				\State \textbf{Parameters:} $\mu, \sigma, \phi, \Delta, \kappa, \psi, num_{bins}, T_{\text{end}}, n_{steps}, {p_i}, N_{\text{sims}}$
				\Statex \textit{\# Initialization:} $G(p_i) \sim \mathcal{N}(\mu, \sigma, p_i)$
				\Statex \textit{\# Reaction Definitions}
				\ForAll{$i \in [1, num_{bins}]$}
				\State $R_{\text{create}, X_i}: \text{pool} \xrightarrow{\phi G(p_i)} X_i, \quad X \in \{A, B\}$
				\State $R_{\text{destroy}, X_i}: X_i \xrightarrow{\Delta n_{X_i}} \text{pool}$
				\EndFor
				\ForAll{$A_i, B_j$ with $p_j \geq p_i$}
				\State $R_{\text{int}, A_i B_j}: A_i + B_j \xrightarrow{\kappa e^{\psi(p_j - p_i)} n_{A_i} n_{B_j}} \text{pool}$
				\EndFor
				\Statex \textit{\# Simulation}
				\State Initialize system state (Gaussian sampling)
				\State Run Gillespie algorithm up to $T_{\text{end}}$ for $N_{\text{sims}}$ trajectories
				\State Average trajectories to obtain $N_A(t)$ and $N_B(t)$
				\State For each trajectory, average the last $n_{steps}$ time points, then average across trajectories to obtain $n_a(p)$ and $n_b(p)$
			\end{algorithmic}
		\end{algorithm}

\bibliography{PRE}

\end{document}